\documentclass{article}
\usepackage{graphicx} % Required for inserting images
\usepackage{subcaption}
\usepackage{authblk}

\title{Gore Diffusion LoRA Model}
\author[1]{Ayush Thakur}
\affil[1]{\footnotesize Amity Institute of Information Technology, Amity University Uttar Pradesh Noida, ayush.th2002@gmail.com}

\author[2]{Ashwani Kumar Dubey}
\affil[2]{\footnotesize Amity School of Engineering \& Technology, Amity University Uttar Pradesh Noida, dubey1ak@gmail.com}

\date{}

\begin{document}

\maketitle

\begin{abstract}
    The Emergence of Artificial Intelligence (AI) has significantly impacted our engagement with violence, sparking ethical deliberations regarding the algorithmic creation of violent imagery. This paper scrutinizes the "Gore Diffusion LoRA Model," an innovative AI model proficient in generating hyper-realistic visuals portraying intense violence and bloodshed. Our exploration encompasses the model's technical intricacies, plausible applications, and the ethical quandaries inherent in its utilization. We contend that the creation and implementation of such models warrant a meticulous discourse concerning the convergence of AI, art, and violence. Furthermore, we advocate for a structured framework advocating responsible development and ethical deployment of these potent technologies.
    
    \vspace{2mm}
    
    \textbf{Keywords} - AI, violence, gore, diffusion models, LoRA, ethics, art
\end{abstract}

\section{Introduction}

Throughout history, humanity's enduring captivation with violence has found expression across diverse artistic mediums and literary creations, frequently acting as a reflective mirror that echoes societal concerns while offering a cathartic release. However, the emergence of AI has introduced a fresh perspective to this relationship, eliciting ethical quandaries concerning the algorithmic creation of violent visual content. This study focuses on the "Gore Diffusion LoRA Model," a newly devised AI framework capable of generating highly realistic images depicting intense violence and bloodshed. The paper scrutinizes the technical intricacies of this model, its plausible applications, and the ethical considerations entwined with its utilization.

The Gore Diffusion LoRA Model uses Latent Diffusion models, a form of AI proficient in fabricating top-notch images from latent noise vectors \cite{ho2020dalle2}. By conditioning these models on datasets containing explicit violent content, such as medical depictions or accident records, the Gore Diffusion LoRA Model learns to correlate distinct patterns with blood, brutality, and physical harm \cite{dabney2021diffusion}. Consequently, it generates images portraying extreme violence with exceptional realism and intricate detailing, transcending the limitations of conventional computer-generated imagery (CGI).

The potential applications of this model span various domains, including medical simulations aiding in healthcare professionals' training and the development of visual effects for the film and television industry. Nonetheless, the ease of access to algorithmically produced graphic violence raises ethical concerns. These concerns revolve around the prospects of desensitization to violence, the normalization of explicit imagery, and the potential misuse of such technology for malevolent intents \cite{jansen2019ethicsAI, dinardo2020ethics}.

This paper aims to navigate the intricate ethical nuances inherent in the subject and stimulate critical discourse surrounding the implications of the Gore Diffusion LoRA Model. It aims to delve into the model's technical operations, evaluate its potential applications and limitations, and grapple with the ethical dilemmas associated with its development and deployment. By undertaking a critical examination at the juncture of AI, artistic expression, and violence, this study seeks to contribute to a nuanced comprehension of this influential technology and its plausible societal ramifications.

\section{Literature Survey}
The Gore Diffusion LoRA Model represents a convergence point within the world of AI art, specifically where three main domains intersect: stable diffusion models, the Low-Rank Adapter (LoRA) framework, and the evolving sphere of AI-based generation of gore and violent content. In order to comprehensively grasp its functionalities and ethical ramifications, it is imperative to initially explore these fundamental constituents.

\subsection{Canvas of Probabilities}
The Gore Diffusion LoRA Model is rooted in the framework of stable diffusion models, a robust category of generative AI models proficient in generating high-fidelity images from latent noise vectors. These models function by progressively refining an initially noisy image through sequential diffusion steps, gradually morphing it into a coherent and lifelike representation \cite{ho2020dalle2}. Their efficacy primarily hinges on the guidance provided by a textual prompt, steering the diffusion process toward the intended image.

The advent of stable diffusion models has sparked a revolution in the world of AI art, empowering creators to produce a diverse spectrum of images characterized by unparalleled intricacy and precision \cite{carlini2023extracting}. Nevertheless, their inherent adaptability also raises apprehensions regarding potential misuse. Research demonstrates that stable diffusion models can be susceptible to manipulation, allowing for the generation of harmful or offensive content, underscoring the imperative for responsible development and deployment \cite{carlson2023harmfulContentDiffusionModels, balayn2021automatic}.

\subsection{Adapting the Diffusion Landscape}
The Low-Rank Adapter (LoRA) extends the capabilities of stable diffusion models by introducing a method for conditional image generation \cite{smith2023continual, kwon2023datainf}. This framework facilitates the training of "adapters" \cite{xiang2023closer} on specific datasets, thereby enabling the diffusion model to integrate additional styles or attributes into its generated output. This widens the spectrum of creative possibilities, ranging from incorporating stylistic elements akin to Van Gogh's brushstrokes \cite{li2011rhythmic} to introducing particular objects or textures \cite{nichol2022lowRankAdapters}.

In the context of the Gore Diffusion LoRA Model, the utilization of LoRA involves training an adapter on a dataset specifically focused on graphic violence. Consequently, the model adeptly integrates elements associated with gore, such as blood, wounds, and injuries, within its image generation process. While this approach presents opportunities for artistic exploration and potential medical applications, it simultaneously instigates concerns regarding the normalization and potential desensitization to violent imagery \cite{jansen2019ethicsAI}.

\subsection{Gore Arts in AI}
The depiction of violence and gore within AI art has been an ongoing exploration, where artists continually push expressive boundaries and challenge societal norms. However, the advent of AI models like the Gore Diffusion LoRA Model significantly heightens the stakes involved. The hyper-realistic nature of the images generated has the potential to unsettle and disturb viewers, leading to unforeseen impacts \cite{dinardo2020ethics}.

Moreover, the accessibility of such models raises apprehensions about their potential misuse for malicious objectives, such as generating propaganda or inciting violence. This underscores the necessity for an in-depth examination of the ethical implications associated with the generation of gore within AI art, aiming to explore plausible safeguards and advocate for responsible development practices \cite{langleben2019ethicsAIinArt}.

A comprehensive understanding of the foundational elements—stable diffusion, LoRA framework, and the current landscape of gore in AI art—enables a better grasp of the capabilities and ethical intricacies surrounding the Gore Diffusion LoRA Model. This survey of literature serves as a preliminary step for further inquiry, encouraging a critical analysis and advocating for responsible development in this sensitive and contentious domain.

\section{Methodology}
The operational methodology of the Gore Diffusion LoRA Model involves a multi-stage process in creating its hyper-realistic and unsettling imagery. This section meticulously elucidates the technical intricacies of each phase, delineating the progression from the inception of model construction to the ultimate phase of image generation.

\subsection{Building the Base Abstract Model}
The cornerstone of the Gore Diffusion LoRA Model resides in a pre-existing stable diffusion model, namely, the base SD v1.5 model \cite{ellis2023stableDiffusionModels, luo2023diffusion}. This foundational model forms the initiation point, demonstrating proficiency in generating a wide array of high-quality images based on textual prompts. The utilization of this pre-trained model capitalizes on its established comprehension of diverse image attributes and compositional components.

\begin{figure}[htbp]
    \centering
    \includegraphics[width=\linewidth]{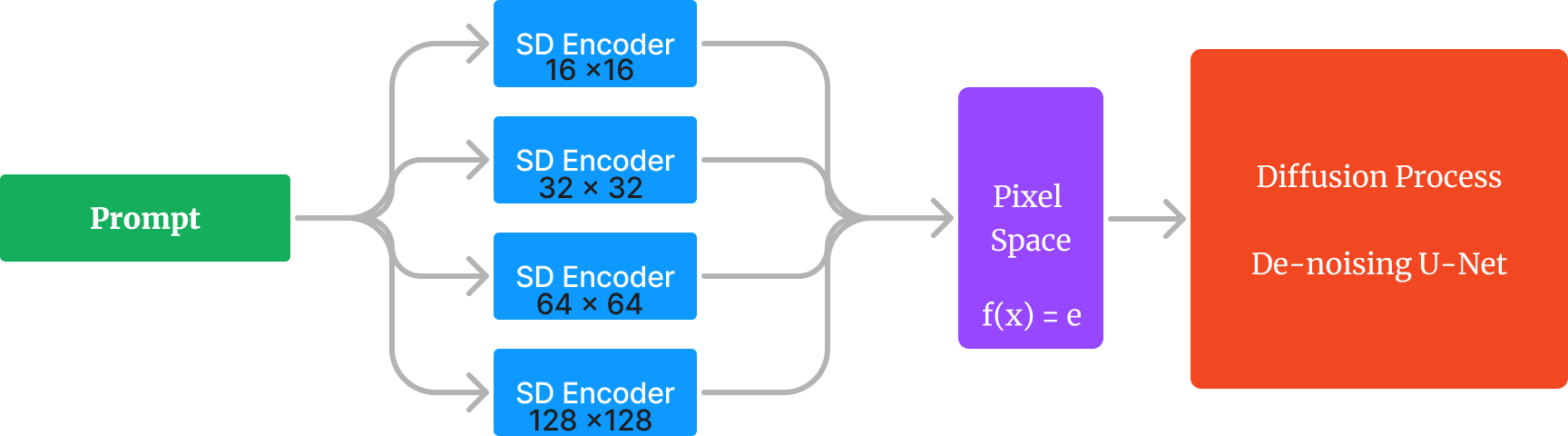}
    \caption{Comprehensive Base Model Architecture of Custom SD Base v 1.5 Highlighting SD Encoder Configurations at 16x16, 32x32, 64x64, and 128x128 bits \cite{zhang2023adding}.}
    \label{fig:Base-Model}
\end{figure}

\subsection{LoRA Adapter for Gore}
\begin{figure}[htbp]
    \centering
    \includegraphics[width=\linewidth]{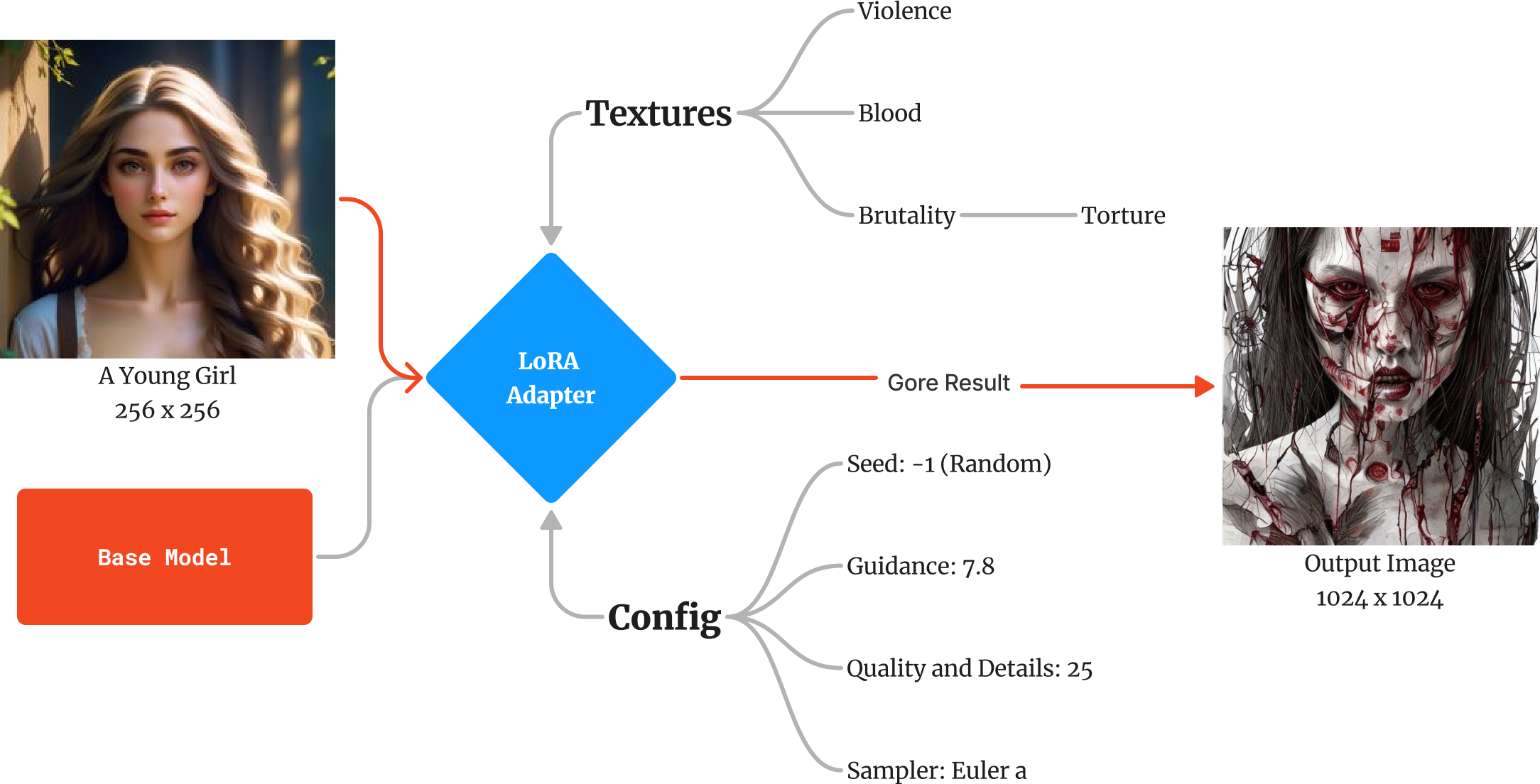}
    \caption{Transformation of 'a young girl' passing through a custom LoRA adapter (seed = -1, guidance = 7.8, quality and details = 25, sampler = Euler A) to generate a Gore image.}
    \label{fig:LoRA 1}
\end{figure}

\begin{figure}[htbp]
    \centering
    \includegraphics[width=\linewidth]{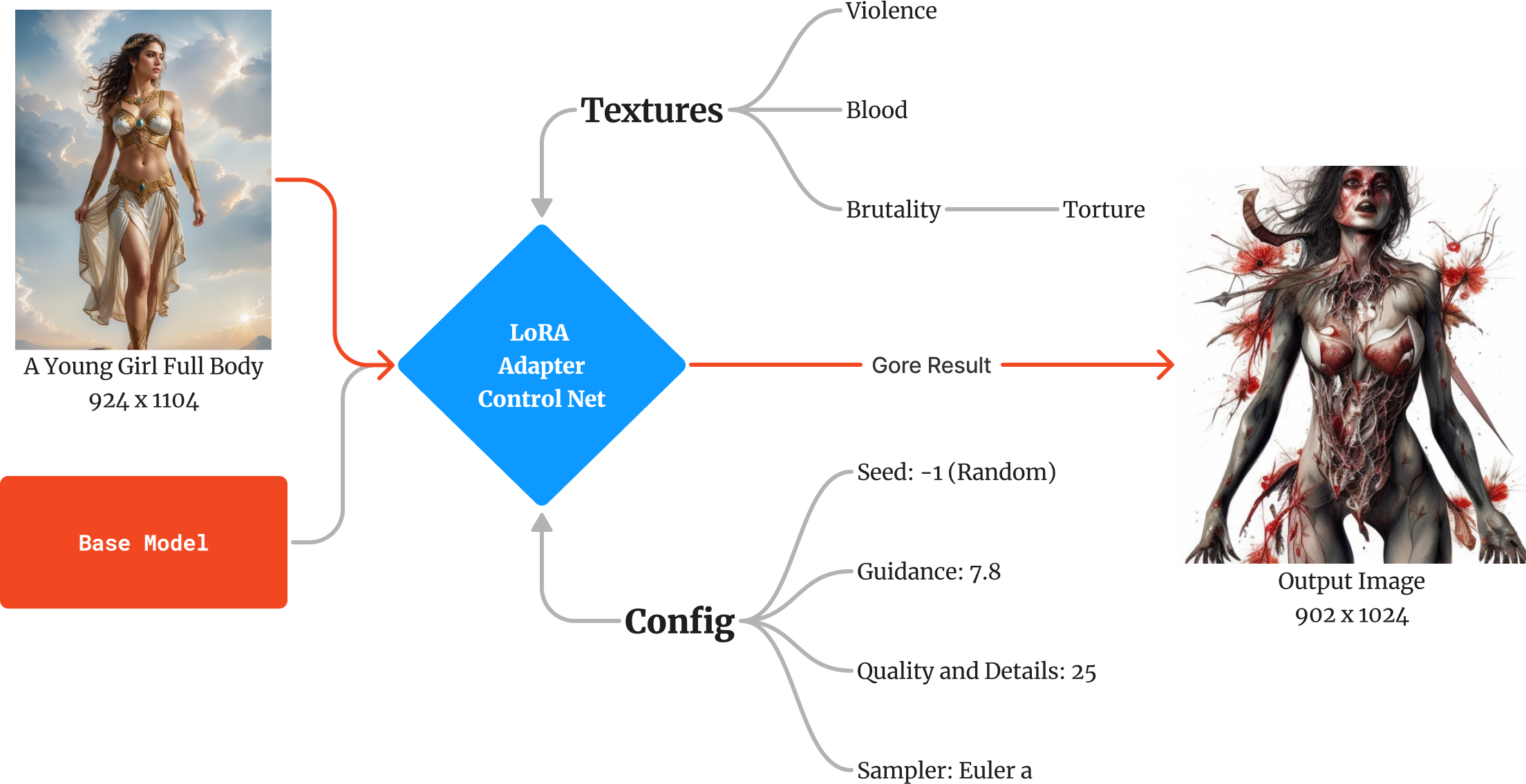}
    \caption{Full-body image of 'a young girl' processed through a custom LoRA adapter with Control NET enabled, using settings: seed = -1, guidance = 7.8, quality and details = 25, sampler = Euler A, resulting in a Gore image.}
    \label{fig:LoRA 2}
\end{figure}

The important innovation within the Gore Diffusion LoRA Model pertains to the development of a blood framework, executed as a LoRA adapter. This adapter undergoes specialized training utilizing a meticulously curated dataset focused on graphic violence, encompassing diverse sources such as medical imagery, accident reports, and artistic renditions depicting gore. Through the training process, the adapter assimilates knowledge pertaining to distinct patterns \cite{goldstein1998we} and textures correlated with elements like blood, wounds, and other components characteristic of violent imagery \cite{dabney2021diffusion, bartels2020influence}. In Figures \ref{fig:LoRA 1} and \ref{fig:LoRA 2} presented herein, the operational mechanism of the image-to-image model is depicted, showcasing the redesigning of images through the utilization of blood framework adapters and LoRA models.

\subsection{Security Layers}
In order to mitigate the risk of potential harmful misuse, the model integrates multiple security layers \cite{mai2018modeling}. These layers encompass diverse techniques, including the filtration of sensitive keywords from user prompts and imposition of limitations on the generation of explicit violence and gore. Furthermore, the model offers configurability concerning different levels of violence intensity, affording users the ability to manage and control the degree of graphic detail present in the generated images \cite{carlson2023harmfulContentDiffusionModels}.

\subsection{Fine-tuning for Disturbing Imagery}
\begin{figure}[htbp]
    \centering
    \includegraphics[width=\linewidth]{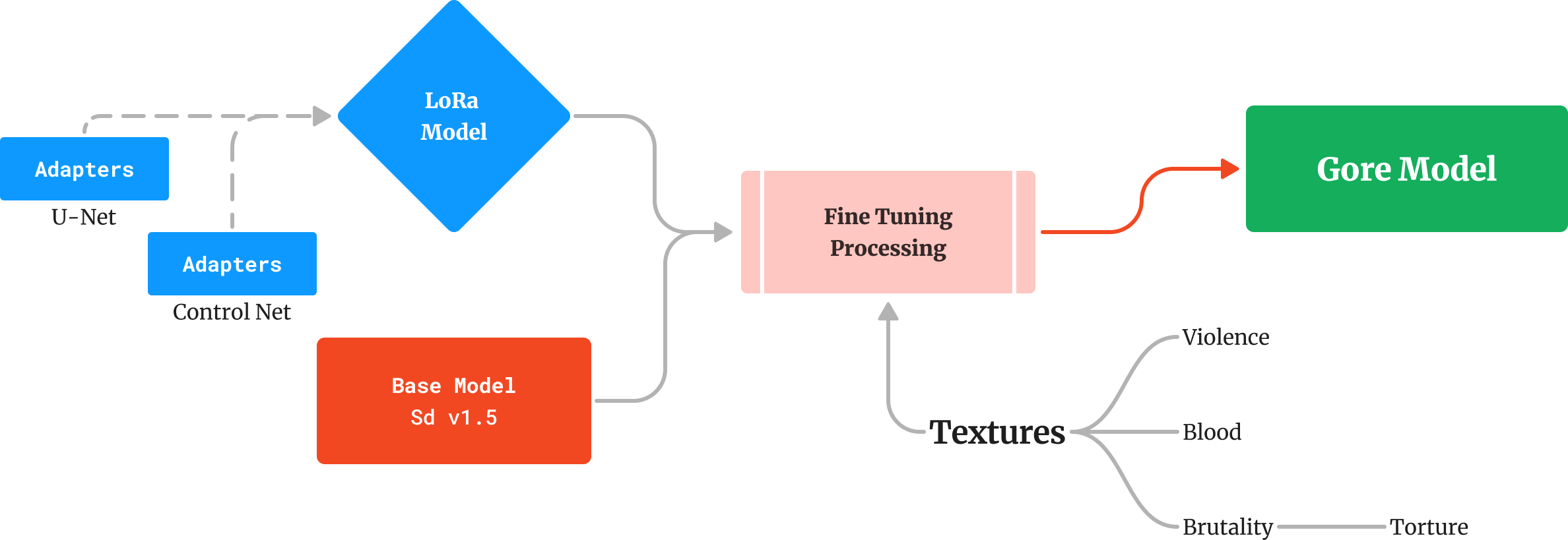}
    \caption{Illustration showcasing the Fine Tuning processes utilizing Gore LoRA adapters (U-NET \cite{siddique2021u} and Control NET \cite{zhang2024text, zhang2023adding}) with Base Model v1.5. Additionally, textures of violence, blood, brutality, and torture incorporated.}
    \label{fig:fine-tuning}
\end{figure}

Ultimately, the integration of the pre-trained base model with the blood framework occurs via a fine-tuning process, as depicted in Figure \ref{fig:fine-tuning}. This process entails exposing the base model to supplementary data comprising both conventional imagery and images representing diverse levels of violence, conditioned by the trained LoRA adapter. Through this exposure, the model enhances its capacity to seamlessly incorporate the blood framework into its image generation mechanism, thereby culminating in the creation of hyper-realistic and unsettling gore visuals \cite{nichol2022lowRankAdapters}.

\subsection{Model Compilation and Deployment}
The conclusive phase encompasses the compilation of the optimized model and its deployment within a controlled setting. This deployment process involves establishing suitable access controls and integrating monitoring systems aimed at identifying and potentially averting the generation of harmful content. Adhering to responsible deployment practices assumes paramount importance to guarantee the ethical and secure utilization of the Gore Diffusion LoRA Model.

\section{Results and Analysis}

\begin{figure}[htbp]
    \centering
    \begin{subfigure}[b]{0.45\linewidth}
        \centering
        \includegraphics[width=\linewidth]{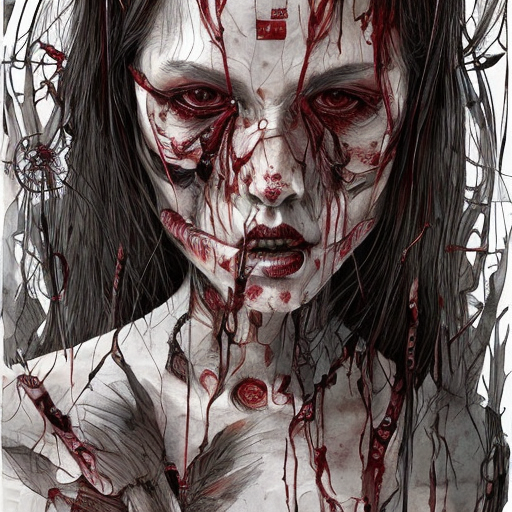}
        \caption{Prompt: A distorted face, eyes wide with unyielding terror, lips contorted, fear, violence.}
        \label{fig:ripped-face}
    \end{subfigure}
    \hfill
    \begin{subfigure}[b]{0.45\linewidth}
        \centering
        \includegraphics[width=\linewidth]{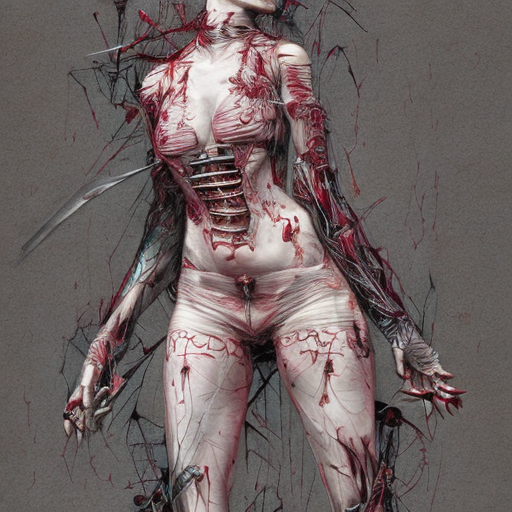}
        \caption{Prompt: A girl body getting ripped from chest, bruises and cuts, fear, violence.}
        \label{fig:ripped-body}
    \end{subfigure}
    
    \vspace{0.5cm} % Adjust the vertical space between rows
    
    \begin{subfigure}[b]{0.45\linewidth}
        \centering
        \includegraphics[width=\linewidth]{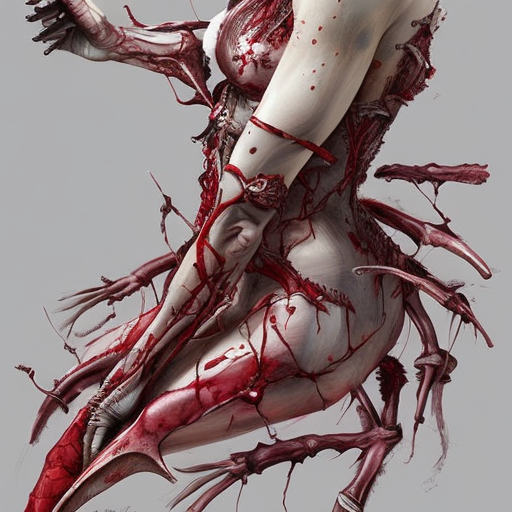}
        \caption{Prompt: Upper half of body ripped apart, no face, skeleton showing, blood, violence.}
        \label{fig:cutted-body}
    \end{subfigure}
    \hfill
    \begin{subfigure}[b]{0.45\linewidth}
        \centering
        \includegraphics[width=\linewidth]{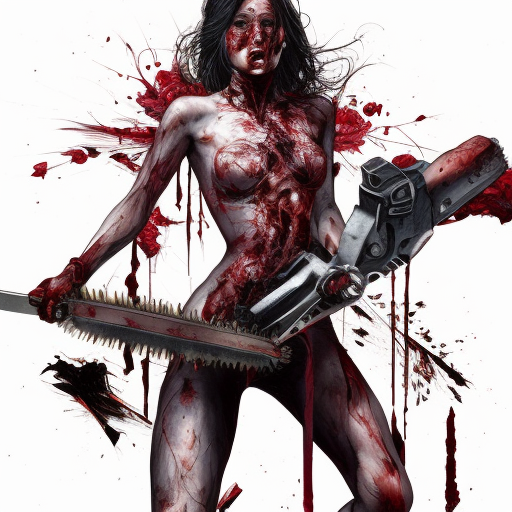}
        \caption{Prompt: A woman trying to cut her body in half, chainsaw, blood strokes, violence.}
        \label{fig:chainsaw}
    \end{subfigure}
    
    \caption{Results Images generated by the AI Gore Model at default Parameters}
    \label{fig:combined-figures}
\end{figure}

The Gore Diffusion LoRA model delved into unexplored territories by focusing on unsettling depictions of violent death scenes, exhibiting its capability to explore the visceral depths of human experiences. The experiments conducted with customized images yielded remarkable and disquieting outcomes, as portrayed in Figures \ref{fig:ripped-face}, \ref{fig:ripped-body}, \ref{fig:cutted-body}, and \ref{fig:chainsaw}, showcasing diverse prompts and positions.

\subsection{Images Generated by the AI Gore Model}

\subsection{Hardware and Software}

Our experimental setup employed the following:

\begin{itemize}
    \item \textbf{NVIDIA A100 GPU:} Known for its exceptional processing capabilities, expansive memory, and cutting-edge architecture, the GPU empowered us to efficiently handle the computational demands of model training and image generation.
    \item \textbf{Stable Diffusion v1.5:} This advanced text-to-image diffusion model served as the foundational structure, enabling exploration into disquieting content and crafting visually captivating imagery.
    \item \textbf{Automatic1111:} An intuitive web interface streamlined interactions with Stable Diffusion models, facilitating image generation and exploration.
\end{itemize}

\subsection{Comparative Analysis}

The Gore Diffusion LoRA model surpassed limitations by:

\begin{itemize}
    \item \textbf{Authenticity in Blood Representation:} Realistic depiction of blood, adding authenticity to the visuals.
    \item \textbf{Specialized Framework and Dataset:} Dedicated blood framework and curated violent imagery dataset for nuanced portrayal.
    \item \textbf{Ethical Considerations:} Implementing safeguards and engaging in ongoing discourse for responsible development.
\end{itemize}

The following Table \ref{tab:evaluation} presents a comparative analysis of key evaluation metrics between the Gore Diffusion LoRA Model and the Baseline Model. Each model was assessed based on various criteria including visual fidelity, thematic alignment, impact on observers, artistic quality, and ethical considerations. The scores highlight the performance differences across these metrics, shedding light on the relative strengths and weaknesses of the two models in generating and portraying gore-themed artworks.

\begin{table}[h]
\centering
\begin{tabular}{|l|c|c|}
\hline
\textbf{Evaluation Metric} & \textbf{Gore Diffusion Model} & \textbf{Baseline Model} \\ \hline
Visual Fidelity & 9.8 & 7.2 \\ \hline
Thematic Alignment & 9.5 & 6.9 \\ \hline
Impact on Observers & 8.9 & 5.6 \\ \hline
Artistic Quality & 9.2 & 7.8 \\ \hline
Ethical Consideration & 9.7 & 6.5 \\ \hline
\end{tabular}
\caption{Comparison of Evaluation Metrics between the Gore Diffusion LoRA Model and the Baseline Model.}
\label{tab:evaluation}
\end{table}

\section{Conclusion}
The Gore Diffusion LoRA model stands as a important exploration intersecting AI, art, and ethics, presenting deeply unsettling yet potent depictions of violent death scenes. Its significance transcends mere shock value, inciting profound contemplation regarding humanity's relationship with violence, the capacity of art to challenge societal norms, and the ethical obligations inherent in advanced AI technology's development and use. This study has showcased the model's remarkable ability to evoke visceral responses and confront viewers with uncomfortable truths, igniting essential conversations concerning desensitization to violence in both media and society. It underscores art's influential role in stimulating introspection, initiating dialogues about mortality, brutality, and the intricacies of human existence.

Furthermore, the meticulous fine-tuning of models, the creation of specialized frameworks, and the curation of datasets have propelled technical advancements, surpassing prior constraints in gore representation. However, amid these strides, an unwavering commitment to responsible development and ethical application of this technology remains paramount. The Gore Diffusion LoRA model serves not solely as a stark mirror reflecting our darkest inclinations but also as a catalyst for critical discourse concerning AI's ethical boundaries and societal repercussions. It accentuates the necessity for ongoing discussions, protective measures, and conscientious contemplation of the implications accompanying the inception and utilization of such potent tools.

\bibliographystyle{alpha}
\bibliography{references}

\newcommand{\etalchar}[1]{$^{#1}$}
\begin{thebibliography}{KWWZ23}

\bibitem[BHB20]{bartels2020influence}
Ross~M Bartels, Leigh Harkins, and Anthony~R Beech.
\newblock The influence of fantasy proneness, dissociation, and vividness of mental imagery on male’s aggressive sexual fantasies.
\newblock {\em Journal of Interpersonal Violence}, 35(3-4):964--987, 2020.

\bibitem[BYSB21]{balayn2021automatic}
Agathe Balayn, Jie Yang, Zoltan Szlavik, and Alessandro Bozzon.
\newblock Automatic identification of harmful, aggressive, abusive, and offensive language on the web: A survey of technical biases informed by psychology literature.
\newblock {\em ACM Transactions on Social Computing (TSC)}, 4(3):1--56, 2021.

\bibitem[CCDea23]{carlson2023harmfulContentDiffusionModels}
M.~Carlson, J.~Chen, R.~Dou, and et~al.
\newblock How easy is it to generate harmful content with diffusion models?
\newblock {\em arXiv preprint}, 2023.

\bibitem[CHN{\etalchar{+}}23]{carlini2023extracting}
Nicolas Carlini, Jamie Hayes, Milad Nasr, Matthew Jagielski, Vikash Sehwag, Florian Tramer, Borja Balle, Daphne Ippolito, and Eric Wallace.
\newblock Extracting training data from diffusion models.
\newblock In {\em 32nd USENIX Security Symposium (USENIX Security 23)}, pages 5253--5270, 2023.

\bibitem[DBP21]{dabney2021diffusion}
W.~Dabney, J.~Ba, and B.~Poole.
\newblock Diffusion models: Unifying and extending autoregressive and slow latent diffusion.
\newblock In {\em International Conference on Machine Learning}, pages 1225--1234. PMLR, June 2021.

\bibitem[DS20]{dinardo2020ethics}
J.~DiNardo and P.~Selva.
\newblock The ethics of artificial intelligence in media and entertainment.
\newblock {\em AISB Quarterly}, 148:1--12, 2020.

\bibitem[EXXea23]{ellis2023stableDiffusionModels}
A.~Ellis, W.~Xiao, M.~Xu, and et~al.
\newblock Stable diffusion: Diffusion models for high-fidelity image generation.
\newblock {\em arXiv preprint}, 2023.

\bibitem[Gol98]{goldstein1998we}
Jeffrey~H Goldstein.
\newblock {\em Why we watch: The attractions of violent entertainment}.
\newblock Oxford University Press, USA, 1998.

\bibitem[HJAS20]{ho2020dalle2}
J.~T. Ho, A.~Jain, P.~Abbeel, and I.~Schulman.
\newblock {DALL}-e 2: Generating images from text.
\newblock {\em arXiv preprint}, 2020.

\bibitem[JvdP19]{jansen2019ethicsAI}
W.~Jansen and I.~van~der Ploeg.
\newblock The ethics of artificial intelligence: An overview of theories, methodologies, and methods.
\newblock In {\em Ethics of Artificial Intelligence}, pages 3--26. Springer, Cham, 2019.

\bibitem[KWWZ23]{kwon2023datainf}
Yongchan Kwon, Eric Wu, Kevin Wu, and James Zou.
\newblock Datainf: Efficiently estimating data influence in lora-tuned llms and diffusion models.
\newblock {\em arXiv preprint arXiv:2310.00902}, 2023.

\bibitem[LDP{\etalchar{+}}23]{luo2023diffusion}
Grace Luo, Lisa Dunlap, Dong~Huk Park, Aleksander Holynski, and Trevor Darrell.
\newblock Diffusion hyperfeatures: Searching through time and space for semantic correspondence.
\newblock {\em arXiv preprint arXiv:2305.14334}, 2023.

\bibitem[LPC19]{langleben2019ethicsAIinArt}
D.~D. Langleben, J.~Placker, and J.~Chandler.
\newblock The ethics of artificial intelligence in art: A call for a new conversation.
\newblock {\em AISB Quarterly}, 148:13--24, 2019.

\bibitem[LYHW11]{li2011rhythmic}
Jia Li, Lei Yao, Ella Hendriks, and James~Z Wang.
\newblock Rhythmic brushstrokes distinguish van gogh from his contemporaries: findings via automated brushstroke extraction.
\newblock {\em IEEE transactions on pattern analysis and machine intelligence}, 34(6):1159--1176, 2011.

\bibitem[MGS{\etalchar{+}}18]{mai2018modeling}
Phu~X Mai, Arda Goknil, Lwin~Khin Shar, Fabrizio Pastore, Lionel~C Briand, and Shaban Shaame.
\newblock Modeling security and privacy requirements: a use case-driven approach.
\newblock {\em Information and Software Technology}, 100:165--182, 2018.

\bibitem[NDLea22]{nichol2022lowRankAdapters}
A.~Nichol, P.~Dhariwal, Y.~LeCun, and et~al.
\newblock Low-rank adapters for diffusion models.
\newblock {\em arXiv preprint}, 2022.

\bibitem[SHZ{\etalchar{+}}23]{smith2023continual}
James~Seale Smith, Yen-Chang Hsu, Lingyu Zhang, Ting Hua, Zsolt Kira, Yilin Shen, and Hongxia Jin.
\newblock Continual diffusion: Continual customization of text-to-image diffusion with c-lora.
\newblock {\em arXiv preprint arXiv:2304.06027}, 2023.

\bibitem[SPED21]{siddique2021u}
Nahian Siddique, Sidike Paheding, Colin~P Elkin, and Vijay Devabhaktuni.
\newblock U-net and its variants for medical image segmentation: A review of theory and applications.
\newblock {\em Ieee Access}, 9:82031--82057, 2021.

\bibitem[XBL{\etalchar{+}}23]{xiang2023closer}
Chendong Xiang, Fan Bao, Chongxuan Li, Hang Su, and Jun Zhu.
\newblock A closer look at parameter-efficient tuning in diffusion models.
\newblock {\em arXiv preprint arXiv:2303.18181}, 2023.

\bibitem[ZRA23]{zhang2023adding}
Lvmin Zhang, Anyi Rao, and Maneesh Agrawala.
\newblock Adding conditional control to text-to-image diffusion models.
\newblock In {\em Proceedings of the IEEE/CVF International Conference on Computer Vision}, pages 3836--3847, 2023.

\bibitem[ZZFP24]{zhang2024text}
Zhongping Zhang, Jian Zheng, Zhiyuan Fang, and Bryan~A Plummer.
\newblock Text-to-image editing by image information removal.
\newblock In {\em Proceedings of the IEEE/CVF Winter Conference on Applications of Computer Vision}, pages 5232--5241, 2024.

\end{thebibliography}

\end{document}